**Ecology (accepted)**

# Seed mass diversity along resource gradients: the role of allometric growth rate and size-asymmetric competition

Niv DeMalach[1]* (nivdemalach@gmail.com)

Ronen Kadmon[1] (kadmon@mail.huji.ac.il)

[1]Department of Ecology, Evolution and Behavior, The Hebrew University of  Jerusalem, Givat Ram, Jerusalem, 91904, Israel

* Corresponding author: Tel. 972-2-6584659, Fax: 972-2-6858711

**Running title**: Seed mass diversity along resource gradients

**Key words:** seed size, size-number trade-off, competition-colonization trade-off, disturbance, functional traits, functional diversity, productivity, community weighted mean, community weighted variance, tolerance-fecundity trade-off,

**ABSTRACT**

The large variation in seed mass among species inspired a vast array of theoretical and empirical research attempting to explain this variation. So far, seed mass variation was investigated by two classes of studies: one class focuses on species varying in seed mass *within* communities, while the second focuses on variation *between* communities, most often with respect to resource gradients. Here, we develop a model capable of simultaneously explaining variation in seed mass *within* and *between* communities. The model describes resource competition (for both soil and light resources) in annual communities and incorporates two fundamental aspects: light asymmetry (higher light acquisition per unit biomass for larger individuals) and growth allometry (negative dependency of relative growth rate on plant biomass). Results show that both factors are critical in determining patterns of seed mass variation. In general, growth allometry increases the reproductive success of small-seeded species while light asymmetry increases the reproductive success of large-seeded species. Increasing availability of soil resources increases light competition, thereby increasing the reproductive success of large-seeded species and ultimately the community (weighted) mean seed mass. An unexpected prediction of the model is that maximum variation in community seed mass (a measure of functional diversity) occurs under intermediate levels of soil resources. Extensions of the model incorporating size-dependent seed survival and disturbance also show patterns consistent with empirical observations. These overall results suggest that the mechanisms captured by the model are important in determining patterns of species and functional diversity.

**INTRODUCTION**

The large variation in seed mass within and between plant communities has fascinated ecologists for decades (Salisbury 1942, Harper et al. 1970, Baker 1972, Moles et al. 2007). Explaining this variation is crucial for understanding the structure and diversity of plant communities because seed mass is an essential component of plant fitness, affecting competitive ability (Leishman 2001, Turnbull et al. 2004), tolerance to environmental stressors (Baker 1972, Leishman and Westoby 1994, Osunkoya et al. 1994, Saverimuttu and Westoby 1996), seed predation (Thompson 1987), dispersal (Greene and Johnson 1993) and dormancy (Harel et al. 2011). Moreover, since the amount of resources available for reproduction is never unlimited, there is an inherent trade-off between seed mass and seed number, and any investment in seed mass must come at the expense of offspring number (Smith and Fretwell 1974). This inherent trade-off has important evolutionary and ecological implications (e.g., Rees and Westoby 1997, Eriksson and Eriksson 1998, Coomes et al. 2002, Falster et al. 2008, Ben-Hur et al. 2012).

Previous attempts to explain variation in seed mass in plant communities have focused on two distinct questions: (1) what mechanisms are capable of maintaining seed mass diversity *within* a community (e.g., Rees and Westoby 1997, Eriksson and Eriksson 1998, Leishman and Murray 2001, Coomes et al. 2002, Franzen 2004, Turnbull et al. 2008, Cornwell and Ackerly 2009, Muller-Landau 2010, Ben-Hur et al. 2012, Lonnberg and Eriksson 2012, D'Andrea et al. 2013); and (2) what mechanisms generate variation in seed mass *between* communities (e.g., Fernandez Ales et al. 1993, Pakeman et al. 2008, Schamp et al. 2008, Metz et al. 2010, Viard-Cretat et al. 2011, Bernard-Verdier et al. 2012, Carmona et al. 2015, Lhotsky et al. 2016).

Although these components of seed mass variation are strongly interconnected, mechanisms maintaining variation within communities have usually been studied independently of mechanisms accounting for between-community variation, and no attempt has been made to develop a theoretical framework capable of simultaneously explaining both aspects of variation.

Here, we present a theoretical framework that attempts to simultaneously explain both phenomena. While the basic mechanisms and processes incorporated in the proposed framework are general, the manner by which we model plant growth and population dynamics were fitted to annual plants due to their simple population dynamics (no generation overlap) and the fact that much of the previous theoretical (e.g., Schwinning and Fox 1995) and empirical (e.g., Turnbull et al. 2004, Ben-Hur et al. 2012, May et al. 2013) studies have focused on annual systems. Our paper is organized in four parts. In the first part, we briefly describe the current models of seed mass variation. In the second part, we develop our theoretical model. In the third part, we use the model as a framework for analyzing the mechanisms underlying variation in seed mass within and between communities. In particular, we focus on the role of resource availability gradients because empirical studies indicate that this factor is a major determinant of seed mass variation in plant communities (e.g., Fernandez Ales et al. 1993, Metz et al. 2010, Pakeman 2011, Bernard-Verdier et al. 2012, Carmona et al. 2015, Lhotsky et al. 2016). Finally, we discuss our theoretical results with respect to previous theoretical and empirical studies of seed mass variation and the potential contribution of our modeling approach to the broader field of 'trait based ecology' (McGill et al. 2006, Shipley et al. 2016).

**Brief description of existing models**

Most previous models attempting to explain within-community variation in seed mass assumed a trade-off between seed mass and seed number, where large seeds have some advantage that compensates for their lower number. Depending on the model, this advantage might be a higher competitive ability (the '*competition-colonization trade-off hypothesis*', Rees & Westoby 1997; Geritz, van der Meijden & Metz 1999; Turnbull *et al.* 2004), or higher tolerance to environmental stress (the '*tolerance-fecundity trade-off hypothesis*', Muller-Landau 2010; Adler *et al.* 2013; D'Andrea, Barabas & Ostling 2013; Haegeman, Sari & Etienne 2014). A fundamental assumption in most of these models is that the total amount of resources available for reproduction is constant among species, creating a trade-off between seed size and seed number (e.g., Smith and Fretwell 1974, Rees and Westoby 1997, Coomes and Grubb 2003, Muller-Landau 2010). While this assumption is a reasonable approximation in some communities, plant species often show considerable differences in both their adult size and reproductive biomass (Aarssen 2005, Falster et al. 2008). Nevertheless, little attention has been paid to the manner by which such differences affect patterns of seed mass variation (though see Falster et al. 2008, Venable and Rees 2009)**.** Another common attribute of these models is the focus on seed fate rather than the biomass growth dynamics that determine seeds production.

The variation in seed mass *between* communities has often been investigated with respect to gradients of soil resources, assuming that soil resource availability (hereafter, 'resource availability') is a major determinant of the fate of small-seeded *vs*. large-seeded species (e.g., Fernandez Ales et al. 1993, Metz et al. 2010, Pakeman et al. 2011, Bernard-Verdier et al. 2012, Carmona et al. 2015, Lhotsky et al. 2016). Although such studies provide clear evidence for variation in seed size along resource gradients, the interpretation of the observed patterns and the conclusions derived from them lack a solid basis due to the lack of theoretical models.

We propose that two fundamental factors that are particularly important for explaining seed mass variation based on resource acquisition and biomass growth, size-asymmetry of light competition (the fact that large individuals receive more light per unit biomass than small individuals), and the allometry of plant growth (the fact that relative growth rate decreases with size, Turnbull et al. 2012). Size asymmetry of light competition (hereafter 'light asymmetry') can be expected to affect patterns of seed mass variation because it increases the competitive advantage of large-seeded species (Tilman 1988, Schwinning and Fox 1995). Growth allometry is expected to influence patterns of seed mass variation because it generates differences in growth rate between large-seeded and small-seeded species (Falster et al. 2008, Turnbull et al. 2012). It can also be expected that the two factors would interact in determining patterns of seed mass variation, because differences in growth rate generate differences in size, thereby affecting size-dependent processes. Yet, although light asymmetry and growth allometry are universal phenomena affecting plant growth, we currently lack a theoretical framework that would allow us to examine their consequences for patterns of seed mass variation.

In this study, we develop a resource competition model that explicitly incorporates differences among species in seed mass, growth allometry, and light asymmetry. We then use our model to investigate the manner by which light asymmetry, growth allometry and resource availability affect the reproductive success of species with varying seed mass, and how such differences are 'translated' into patterns of seed mass variation within and between communities that differ in resource availability.

**THE MODEL**

Our model focuses on annual species that vary in their seed mass. The model describes resource competition (for soil resource and light) and biomass growth of individuals during the growing seasons and their consequences for population dynamics and community structure. According to the model, the biomass ($S$) of individual $i$ in day ($d$ +1) is determined by its biomass in day ($d$),

its maximal relative growth rate ($\mu$, the growth rate per unit biomass [RGR] in the absence of

any resource limitation), and a parameter $p$ indicating the degree to which growth rate is reduced

by resource limitation:

(1)  $S_{i\,(d+1)} = S_{i(d)} + S_{i(d)}{}^{a} \cdot \mu_i \cdot p_{i(d)}$

The exponent ($a$) in the equation introduces growth allometry and ranges from zero to one,

where one leads to size independent RGR and values smaller than one lead to allometric growth

(i.e., a reduction in RGR with increasing size). Note that absolute growth rate increases with size

also under allometric growth. For simplicity we assumed that $\mu$ is equal for all species (but see

Appendix S1 for relaxing this assumption).

We further assume that a single resource limits plant growth in each time step ('Liebig's law of

the minimum'):

(2)  $p_{i(d)} = \min(\frac{r_{\text{soil},i(d)}}{k_{soil}+r_{\text{soil},i(d)}}, \frac{r_{light,i(d)}}{k_{light}+r_{light,i(d)}})$.

This equation is based on a Michaelis-Menten (Monod) growth function, where the value of $p$

ranges from zero (no growth) to one (maximal growth) and $k$ is the half saturation constant. The variables $r_{\text{soil},i(d)}$ and $r_{light,i(d)}$ represent the *net* amount of soil resources and light *per unit* *biomass* of individual $i$ at day ($d$), after taking into account a resource-specific maintenance cost (DeMalach et al. 2016):

(3)   $r_{ji(d)} = \frac{R_{ji(d)}}{S_{i(d)}} - M_{ji}.$

$R_{ji(d)}$ is the amount of resource $j$ (light or soil resource) available *for the whole plant* at day ($d$), $M_{ji}$ is a parameter indicating the maintenance cost of resource $j$ for individual $i$ (*per biomass unit*), and $r_{ji(d)}$ is the net amount of resource $j$ available per unit biomass after taking into account the maintenance cost. If maintenance cost ($M$) exceeds the amount of resource available per unit biomass ($R/S$), growth rate is set to zero (i.e., no negative growth). The model assumes no mortality of plants under negative balance (but see appendix S2 for relaxing this assumption).

It is further assumed that each resource is supplied at some constant rate and that the amount of resource $j$ available for plant $i$ at a given day ($R_{ji(d)}$) is determined by the size-asymmetry of resource exploitation (Schwinning and Weiner 1998, DeMalach et al. 2016):

(4)   $R_{ji(d)} = \bar{R}_j \cdot \frac{S_{i(d)}^{\theta}}{\sum_{i=1}^{n} S_{i(d)}^{\theta}}$

where $\bar{R}_j$ is the supply rate (total amount of resource $j$ supplied to the system at each day), and $\theta$ indicates the level of size asymmetry in resource exploitation. When $\theta = 1$, resource exploitation is size symmetric (each plant gets the same amount of resource *per unit biomass*), whereas as $\theta$ increases, resource exploitation becomes more asymmetric and larger plants get larger amounts of resource per unit biomass than smaller plants. The soil resource was assumed to be size-symmetric in all simulations (i.e., $\theta_{soil} = 1$). The size asymmetry in light acquisition ($\theta_{light}$) was 1 or greater than 1, in order to test its effect on the predictions of the model. Importantly, equation

(4) assumes that all individuals are well mixed and compete for the same pool of resources (i.e., a complete niche overlap).

At the end of the growing season, each individual produces $O$ seeds by allocating some proportion ($\alpha$) of its biomass to reproduction:

(5) $\quad O_i = \alpha \cdot \frac{S_{i(end)}}{S_{i(0)}}$

where $O_i$ is the number of seeds produced per individual plant, $S_{i(end)}$ is the final biomass, $S_{i(0)}$ is biomass of a single seed, and $\alpha$ is the proportion of biomass allocated to reproduction. Importantly, all individuals of a given species are identical in our model, but the biomass at the end of the growing season ($S_{i(end)}$) varies between species, depending on the complex interactions between initial size, allometry, resource availability and size-asymmetry. Hence, unlike many previous models, our model does not assume a trade-off between seed size and seed number.

For simplicity, we assumed that all seeds produced at the end of the growing season are available for germination in the next year (i.e., the model does not incorporate seed dormancy and seed mortality but see appendix S3) and that the initial biomass of a germinant equals its seed mass. Under these assumptions, population size ($N$) of species $k$ in year ($y+1$) is simply determined by seed production of all individuals of that species in the previous year:

(6) $N_{k\,(y+1)} = \sum_{i=1}^{N_{k\,(y)}} O_{i(y)}$

To increase the realism of the model we incorporated demographic stochasticity (hereafter, 'drift') by assuming that actual population size at year ($y + 1$) is drawn from a Poisson distribution (Kalyuzhny and Shnerb 2017) with a mean $(N_{k(y+1)})$. Drift is considered one of the main processes determining community assembly, so any selection processes need to override the constant drift in order to be detected in real world communities (Vellend 2010). Furthermore, recently, it was demonstrated that drift could be important even when competitive difference are strong (Gilbert & Levine 2017). Nonetheless, we also investigated the model predictions without drift (Appendix S4).

**METHODS**

We used our model to investigate the role of growth allometry ($a$), light asymmetry ($\theta_{light}$) and resource availability ($\bar{R}_{soil}$), on the reproductive success and population dynamics of 100 species competing for both light and soil resources. Soil resources were exploited symmetrically (i.e., $\theta_{soil}$ = 1) in all simulations. Each species had a different seed mass ($S_{i0}$) that was drawn from a lognormal distribution with mean = -2 and SD = 1.2. In all simulations, the initial abundance of each species was 10 individuals.

We performed two sets of simulations. In the first set, we investigated the consequences of differences in seed mass for the reproductive success of competing species *within a single growing season* (from seed germination to seed production). In the second, we extended the analysis into simulations of *inter-annual population dynamics*, and tested how the differences in reproductive success are translated into patterns of seed mass variation within and between communities. For simplicity, we assumed that all species were identical in all traits except seed mass (see Table 1 for all parameter values and appendix S1 for relaxing this assumption).

**Simulations focusing on reproductive success (within-year simulations)**

The first set of simulations focused on factors affecting seed mass variation within a community. In these simulations we tested whether and how growth allometry ($a = 0.5, 0.75, 1$) and light asymmetry ($\theta_{light} = 1, 1.15, 1.3$) affect the reproductive success of competing species varying in their seed mass. A relative measure of reproductive success (hereafter RRS) was determined for each species by dividing its seed production (number of seeds produced at the end of the growing season) by the mean seed production per species in the community. Thus, RRS values smaller than one indicate lower than average RRS and vice versa.

We hypothesized that both growth allometry and light asymmetry are important in determining the reproductive success of species in the community, but that the two factors work in opposite directions with growth allometry increasing the relative success of small-seeded species and light asymmetry increasing the relative success of large-seeded species. These simulations were performed under constant environmental conditions ($\bar{R}_{soil} = 800$, $\bar{R}_{light} = 1000$), thereby, focusing on seed mass variation *within* communities.

We then tested how differences in soil resource ($\bar{R}_{soil} = 300, 800, 1200, 2000$) affect the relationship between seed mass and reproductive success of competing species. In all those simulations light levels were constant ($\bar{R}_{light} = 1500$). Following results from the first set of analyses indicating that both growth allometry and light asymmetry affect RRS, these simulations were performed under a factorial design of growth allometry ($a = 1$ and $0.75$) and light asymmetry ($\theta_{light} = 1$ and $1.15$). We expected that increasing levels of soil resources would increase RRS of large-seeded species and decrease RRS of small-seeded species. However, assuming that the competitive advantage of large-seeded species increases with increasing light asymmetry and decreases with increasing growth allometry, we expected that the degree to which increasing resource availability facilitates the RRS of large-seeded species would increase with increasing light asymmetry and decrease with increasing growth allometry.

**Simulations focusing on seed mass variation (inter-annual simulations)**

The previous simulations focused on the dynamics of individual growth within a single growing season and provided information on the manner by which growth allometry, light asymmetry, and resource availability affect the reproductive success of large- vs. small-seeded species in the community. A second set of simulations was performed to investigate the consequences of these size-dependent reproductive responses for the distribution of seed mass among species in the community, and the manner by which this distribution is influenced by differences in resource availability. These simulations were performed under four levels of resource availability ($\bar{R}_{soil} = 300, 800, 1200, 2000$) and over longer time scale (30 generations).

At the end of each simulation, we quantified two basic characteristics of seed mass distribution that are often reported in empirical studies of functional trait variation (Pakeman et al. 2008): the abundance-weighted mean (commonly termed community weighted mean, CWM) and the abundance-weighted variance (community-weighted variance, CWV):

$$CWM = \sum_i^n P_i T_i.$$

$$CWV = \sum_i^n P_i (T_i - MEAN_i)^2$$

$P_i$ is the relative abundance of species ($i$), $T_i$ is the trait value (seed mass in our case) for species ($i$), and $MEAN_i$ is the mean trait value in the community. Both measures were quantified using $\log_{10}$-transformed values of seed mass (Pakeman et al. 2008).

**RESULTS**

**Simulations focusing on reproductive success (within-year simulations)**

The results of simulations focusing on a single year are presented as the relationship between seed mass and relative reproductive success (RRS) of all species in the community with species ranked according to their seed mass (hereafter, 'RRS-seed mass relationship', Figs. 1, 2). Each simulation generates a unique RRS-seed mass relationship that represents a particular parameter regime. The peak of the RRS response represents the value of seed mass that maximizes RRS and can be interpreted as the optimal seed mass 'strategy' under the relevant conditions.

*Effect of light-asymmetry and growth allometry*

Under the simplest scenario where light competition is size-symmetric ($\theta_{light} = 1$) and relative growth rate is isometric ($a = 1$), seed mass does not affect RRS and all species have the same RRS regardless of their seed mass (i.e., a fully neutral community where all species produce equal amount of seeds, Fig. 1a). Increasing light asymmetry under isometric growth ($a = 1$) provides a competitive advantage for large-seeded species and increases their RRS resulting in a monotonic increase of RRS with increasing seed mass (Fig. 1a-c). Still, above some size threshold, all the species are limited by soil resources (they acquire enough light) so a further increase in seed mass does not convey higher fitness (since the soil resource is exploited symmetrically). Increasing allometry (i.e., decreasing $a$) under size-symmetric light competition ($\theta_{light} = 1$) has an opposite effect and increases the RRS of small-seeded species, resulting in a monotonic decline of RRS with increasing seed mass (Fig. 1a, d, g). Under the more realistic scenario of asymmetric light competition ($\theta_{light} > 1$) and allometric growth ($a < 1$), neither force dominates the pattern of variation and RRS may show a unimodal response to variation in seed mass with maximum RRS being obtained at intermediate seed mass (Fig 1f). In such cases, the level of seed mass that maximizes RRS increases with increasing light asymmetry (compare Fig. 1e, f) and decreases with increasing allometry (compare Fig. 1f, i).

Interestingly, the results above contrast the classical models that predict a negative relationship between seed mass and seed number. The model suggests that the relationship between seed mass and RRS (a relative measure for seed number) could vary depending on growth allometry and light asymmetry.

*Effect of soil resource availability*

The effect of soil resource availability ($\bar{R}_{soil}$) on the RRS-seed mass relationship depends on both the competitive regime and the allometry of growth. If competition is size-symmetric and relative growth rate is isometric (Fig. 2a), the RRS-seed mass relationship remains flat (neutral) and is unaffected by soil resource availability. If light competition is size-asymmetric and relative growth rate is isometric (Fig. 2b), increasing resource availability increases the RRS of large-seeded species. If light competition is size-symmetric and relative growth rate is isometric (Fig. 2c), the RRS-seed mass relationship decreases monotonically and increasing resource availability increases the RRS of small-seeded species. If light competition is size-asymmetric and relative growth rate is allometric (Fig. 2d), RRS shows a unimodal response to variation in seed mass and the seed mass that maximizes RRS increases with increasing levels of soil resources. This pattern results from the increase in the relative importance of size-asymmetric light competition (compared with size-symmetric belowground competition) with increasing soil resource levels favoring large seeded species.

**Simulations focusing on seed mass variation (inter-annual simulations)**

The simulations of population dynamics show that differences in RRS have profound effects on the relative abundance of small-seeded *vs*. large-seeded species (Fig. 3). In general, low levels of soil resources lead to communities dominated by small-seeded species (red lines in Fig. 3) and high resource levels lead to dominance of large-seeded species (blue lines). Intermediate resource levels reduce the dominance of both small-seeded and large-seeded species, and promote the persistence of species with intermediate seed mass (green lines).

Figure 4 summarizes the causal links between RRS (Fig. 4a), species abundance (Fig. 4b), and the two measures of seed size distribution (Fig. 4c). As can be expected, the increase in RSS of large-seeded species at high levels of resource availability increases their relative abundance in the community, resulting in a positive correlation between soil resource availability and community weighted mean (CWM) seed mass (Fig. 4c). Interestingly, the corresponding effect on community weighted variance (CWV) is unimodal with maximum variance in seed mass occurring under intermediate levels of resources (Fig. 4c).

**DISCUSSION**

We investigate mechanisms and consequences of seed mass variation within and between communities using a simple model of annual species competing for soil resources and light. A novel feature of our model is an explicit incorporation of growth allometry and size-asymmetric light competition, two fundamental characteristics of plant growth. Our overall results point to the following generalizations:

1. Under a given level of soil resources, growth allometry increases the Relative Reproductive Success (RRS) of small-seeded species, while light asymmetry increases the RRS of large-seeded species.

2. Soil resource availability has a profound effect on RRS of small-seeded *vs*. large-seeded species. Under realistic growth and competitive regimes (i.e., allometric growth and light asymmetry), low resource levels maximize the RRS of small-seeded species, high resource levels maximize the RRS of large-seeded species, and intermediate resource levels maximize the RRS of species with intermediate seed mass.

3. Increasing soil resource availability increases the community weighted mean (CWM) seed mass, and has a unimodal effect on the community weighted variance (CWV).

Below, we discuss the properties of our model and its predictions in comparison with previous theoretical and empirical studies of seed mass variation. We conclude by suggesting that our modeling approach may provide a promising route for improving the theoretical basis of 'trait based ecology' (McGill et al. 2006, Shipley et al. 2016).

**Comparison with previous models**

The main difference between our model and most previous models is that previous models focused on seed survival and colonization while our models focus on *seed production* (as affected by biomass growth). Hence, most early models *assumed* a competitive advantage of large-seeded species and demonstrated that such advantage could explain coexistence of large-seeded and small-seeded species through a competition-colonization trade-off (e.g., Rees and Westoby 1997, Geritz et al. 1999). In contrast, our model describes plant growth and resource acquisition, and the competitive advantage of large- *vs.* small-seeded species *arises* from the fundamental characteristics of plant growth, rather than assumed. The model confirms the assumption that a large seed mass may lead to competitive advantage, but demonstrates that this advantage is far from being universal, and depends on the degree of growth allometry and size-asymmetry of light competition (Fig. 1). This result is important, because growth allometry is a fundamental element of plant growth (Turnbull et al. 2012), size asymmetry is a fundamental property of light competition (Schwinning and Weiner 1998, DeMalach et al. 2017), and the two factors operate in opposite directions in determining the fitness of large- *vs.* small-seeded species (Fig. 1).

Moreover, our results indicate that the balance between the contrasting effects of growth allometry and size asymmetric light competition depends on soil resource availability (Fig. 2). Increasing resource availability increases the relative importance of light competition and therefore increases the RRS of large-seeded species. Thus, our model predicts that such competitive advantage of large seeded species is limited to situations of high soil resources where light is the dominant limiting factor.

The most fundamental difference between our model and competition-colonization trade-off models concerns the mechanisms underlying the advantage of small-seeded species: in competition-colonization models, this advantage results from higher colonization ability, whereas in our model it results from growth allometry. The latter mechanism has a strong empirical support (Paine et al. 2012, Turnbull et al. 2012), and does not require strict competitive hierarchy as assumed by models of competition-colonization trade-off (Coomes and Grubb 2003).

An alternative class of trade-off models attributes the advantage of large-seeded species to higher fitness under environmentally stressful conditions (Muller-Landau 2010, D'Andrea et al. 2013, Haegeman et al. 2014). A key element of these models is spatial heterogeneity in ecological conditions. Analysing the consequences of such heterogeneity is beyond the scope of our work, but in appendix S3 we show that our theoretical framework can easily accommodate a size-dependent modifier of survival that increases the ability of large-seeded species to tolerate low resource levels, as assumed by this class of models.

Most previous models assumed a perfect trade-off between seed mass and seed number, i.e., when two species vary in their seed size, an individual of the larger-seeded species inevitably produces fewer seeds than an individual of the smaller seeded species (e.g., Rees and Westoby 1997, Coomes and Grubb 2003, Muller-Landau 2010). The justification for this trade off stems from the classical model by Smith and Fretwell (1974) that assumed that adult (reproductive) biomass is *equal* for all species. In contrast, with Smith and Fertwell's assumption, in our model species *vary* in their adult biomass i.e., larger seeded species have also larger adult size. Therefore, in our model there is no inevitable trade-off between seed mass and number (the relationship between seed mass and the number of seeds produced per individual varies depending on the model parameters). We suggest that previous models (e.g., Rees and Westoby 1997, Coomes and Grubb 2003) apply to communities where species differ in their seed mass, but have a similar adult biomass, while our model applies for communities in which species co-vary in their seed and adult biomass (Falster et al. 2008, Venable and Rees 2009).

**Comparison with empirical observations**

According to our model, increasing resource availability increases the abundance of large-seeded species, thereby increasing CWM seed mass. In accordance, seed mass often increases with increasing nutrient availability (Fernandez Ales et al. 1993, Manning et al. 2009, Bernard-Verdier et al. 2012, Santini et al. 2017, Dirks et al. 2017, but see Pakeman et al. 2008, Dainese and Sitzia 2013). Seed mass responses to gradients of water availability are variable, ranging from positive (Harel et al. 2011, May et al. 2013) to neutral (Pakeman et al. 2008, Carmona et al. 2015) and even negative (Marteinsdottir and Eriksson 2014). One possible mechanism that may lead to differences in seed mass responses to nutrient *vs*. soil water gradients is high tolerance of large-seeded species to low levels of water availability following germination (e.g., through faster root development, Baker 1972). Furthermore, in large-seeded species, a greater proportion of the seed biomass often remains undeployed, thereby increasing survival during periods of resource stress (Leishman & Westoby 1994). In contrast, our model is based on the simplifying assumption that all the seed reserves are deployed immediately (which may also exaggerate initial size differences). Nonetheless, higher survival of large seeded species can be easily incorporated in our modeling framework by adding a parameter expressing size-dependent survival (Appendix S3). As expected, such extension reduces the magnitude of decrease in CWM seed mass toward low levels of resource availability, and may even lead to an increase in CWM seed mass at low resource levels. Such a U-shaped seed mass response has been proposed in a recent conceptual model (Bergholz et al. 2015), but to the best of our knowledge, has never been observed in natural communities.

It should also be noted that the prediction of increasing seed mass with increasing resource availability is based on the assumption that soil resources are exploited size-symmetrically. While this assumption is a reasonable approximation for nutrients (Schwinning and Weiner 1998), there are reasons to believe that in some cases, competition for water might also be asymmetric (e.g., due to an increase in water availability with soil depth). Such asymmetry can be incorporated by appropriate parameterization of $\theta$ in our model but currently there is no empirical evidence that justifies such parameterization.

Much research on between-community variation in seed mass has concentrated on the effect of grazing and disturbance. An almost universal pattern emerging from such studies is an increase in the relative abundance of small-seeded species with increasing disturbance (Fernandez Ales et al. 1993, Kahmen et al. 2002, Louault et al. 2005, Peco et al. 2005, Pakeman et al. 2008, but see Niu et al. 2010). This pattern is also predicted by our model when disturbance is incorporated (appendix S5).

An interesting pattern predicted by our model is that variation in seed mass (as expressed by community weighted variance) is maximized at intermediate levels of resource availability (Fig. 4c). We interpret this finding as a result of selection operating at either end of this gradient, which narrows the range of possible seed mass strategies at low and high levels of soil resources. Importantly, this finding is not a mere a consequence of the unimodality of the log-normal seed mass distribution since the same pattern appears even under non-unimodal distribution (appendix S6).

**Limitations of the model**

Patterns of seed mass variation are influenced by numerous processes and no model is capable of simultaneously capturing all of these processes. We focused on what we considered as the most fundamental elements of individual biomass growth and population growth. Nonetheless, there are many other processes by which differences in seed mass may influence the performance of individual plants (e.g., seed dispersal [Greene & Johnson 1993], dormancy [Thompson 1987], emergence probability [Ben-Hur et al. 2013] and seed predation [Tilman 1988]), and incorporating such processes in our model can be expected to generate more complex responses and interactions.

Our model involves simplifying assumptions regarding the germination processes (all seeds germinate together), allocation to different organs (no explicit partitioning of root and shoot) and the spatial structure of the population (all individuals are well mixed). It also ignores the fact that seed mass is often correlated (positively or negatively) with other traits that influence individual fitness (e.g., maximal growth rate could be negatively correlated with seed mass, Rüger et al. 2012). Taking into account such trait-trait constraints is a major challenge for studies of functional trait variation (Douma et al. 2012, Santini et al. 2017) and requires much further theoretical and empirical research. In addition, as in every simulation model, conclusions are based on the parameter space used. While we aimed to investigate a wide-range, the infinite parameter space cannot be fully explored and different parameter values may change the conclusions.

It should also be noted that our model describes non-equilibrium dynamics in a closed system (no immigration) with complete niche overlap (no 'stabilizing mechanism' sensu Chesson 2000). In our model, seed mass differences are translated into 'fitness differences' (sensu Chesson 2000) and ultimately differential exclusion rates. In our model, small fitness differences lead to slow exclusion rates, and vice versa. Still, in a long run the best competitor is expected to exclude all the rest regardless of the parameter space. Alternatively, if constant stabilizing mechanisms (or immigration) were incorporated, decreasing fitness differences would increase the equilibrium number of species and vice versa (Chesson 2000). Here, we decided to keep the model as simple as possible without invoking additional mechanisms (e.g., constant immigration, intraspecific density dependence) and focus on growth allometry and size asymmetric light competition. This choice allows simpler interpretation of the results and enhances empirical tests of this model (few experiments last for more than 30 years). Furthermore, some authors argue that stable coexistence is impossible in the real world due to demographic stochasticity (all species will undergo extinction at some point), and therefore models focusing on extinction rates could better explain diversity patterns compared with classical equilibrium models (e.g., Carmel et al. 2017).

In this contribution we focused on annual plants but the main elements of the model (resource competition, light asymmetry, growth allometry, variation in seed mass) are relevant for both annual and perennial life forms (see Falster et al. 2016 for incorporating these elements in forest communities). Nonetheless, the intergenerational competition that characterizes communities dominated by perennial plants and their overlapping generations may lead to deviations from its current predictions (e.g., Bitomsky et al. 2018). Hopefully, future extensions of the model will explore whether and how relaxing various assumptions of the model affect its main predictions.

**Relevance for trait based ecology**

The recent development of 'trait-based ecology' has stimulated a growing number of studies that incorporate measurements of seed mass (among other key functional traits) in analyses of community structure (e.g., Pakeman et al. 2008, Bernard-Verdier et al. 2012, Diaz et al. 2016). However, currently such studies lack a theory for interpreting their results (Adler et al. 2013). Our model can be considered as one step toward reducing this gap. Although seed mass is only one of many important traits, it is an important determinant of plant fitness, particularly in annual communities (Venable and Brown 1988, Levine and Rees 2002, Turnbull et al. 2004). It also has the advantage of being 'soft' (easy to measure), 'hard' (important), and robust to environmental variation (Weiher et al. 1999). Moreover, our finding that seed mass, growth allometry, and size-asymmetry have a crucial role in determining plant reproductive success through their effects on adult size is fully consistent with a global-scale analysis indicating that plant size is a key component of the worldwide distribution of plant functional traits (Diaz et al. 2016). We therefore believe that these three elements (seed mass, growth allometry, and size-asymmetry) should be an integral part of any future theory of trait based ecology.

**ACKNOWLEDGEMENTS**

We thank Daniel Falster, Joseph Wright and an anonymous reviewer for comments on previous versions of this manuscript. We also thank Eyal Ben-Hur for his technical assistance in performing the simulations. The study was supported by the DFG program "HEDGE II", the Israel Science Foundation grant no. 447/15, the Hebrew University Advanced School of Environmental Studies and the Ring Foundation. The authors declare no conflict of interest.

**DATA ACCESSIBILITY**

The Matlab codes for all simulations are available on Figshare:

https://figshare.com/s/4273d61a5fce9fb950cb

**Table** 1: Parameters used in the simulations. We used similar units ('Abstract Resource Units', ARU) for soil and light resources in order to simplify the construction and interpretation of the model. The last parameter ($y_{end}$) is relevant only for simulations of population dynamics. In addition, the population dynamic simulations assume that $\mu$ is a random variable rather than constant (mean=0.05, SD= 0.005)

| Symbol | Description (units) | Value(s) |
|---|---|---|
| $\mu$ | Maximal relative growth rate (fraction) | 0.05 |
| $M\ (soil\backslash light)$ | Maintenance costs (ARU/mass) | 0.15 |
| $a$ | Allometry exponent (unitless) | 0.5,0.75,1 |
| $k\ (soil\backslash light)$ | Half saturation growth constant (ARU/mass) | 1 |
| $\bar{R}_{light}$ | Light availability (supply rate) (ARU) | 1500 |
| $\bar{R}_{soil}$ | Soil availability (supply rate) (ARU) | 300,800,1200,2000 |
| $\theta_{light}$ | Light asymmetry(unitless) | 1,1.15,1.3 |
| $\theta_{soil}$ | Soil asymmetry (unitless) | 1 |
| $\alpha$ | Reproductive allocation (fraction) | 0.5 |
| $n$ | Number of species in the species pool | 100 |
| $N_{initial}$ | Initial number of individuals per species | 10 |
| $d_{end}$ | Length of the growing season (days) | 200 |
| $y_{end}$ | Length of population dynamics (years) | 30 |

**Figure legends**

**Figure 1**. Effects of light asymmetry ($\theta_{light}$) and growth allometry ($a$) on the relationship between relative reproductive success (RRS) and seed mass, under a constant level of soil resources ($\bar{R}_{soil} = 800$). Results are based on a single growing season.

**Figure 2**: Effect of soil resource availability ($\bar{R}_{soil}$) on the relationship between relative reproductive success (RRS) and seed mass, under different levels of light asymmetry ($\theta_{light}$) and growth allometry ($a$). Results are based on a single growing season

**Figure 3**: Population dynamics of species competing for soil and light resources under different levels of soil resource availability ($\bar{R}_{soil}$). Species are classified into three quantiles based on their seed mass: red – small, green – medium, blue – large. Note the logarithmic scale of the y axis (1 was added to all values before transformation to avoid infinity). All simulations started with 10 individuals per species. Other parameters: $a = 0.75, \theta_{light} = 1.3$

**Figure 4**: A summary of the processes by which the effect of seed mass on relative reproductive success (RRS) is translated into patterns of variation in community weighted mean (CW Mean) and community weighted variance (CW Variance) of seed size along resource gradient. a – effect of seed mass on RRS under different levels of soil resource availability ($\bar{R}_{soil}$) in the first year

(i.e., when all species have the same abundance); b – the resulting relationship between seed mass and species abundance after 30 years (from a single implementation of the model). Note the logarithmic scale of the y axis (1 was added to all values before transformation to avoid infinity). c – effects of resource availability on CW Mean and CW Variance of seed size at the end of the simulation (year 30) based on five implementations of the model; circles and error bars represent means and 95% confidence intervals (not shown when smaller than circle size). Other parameters: $a = 0.75$, $\theta_{light} = 1.3$

**Figure 1**

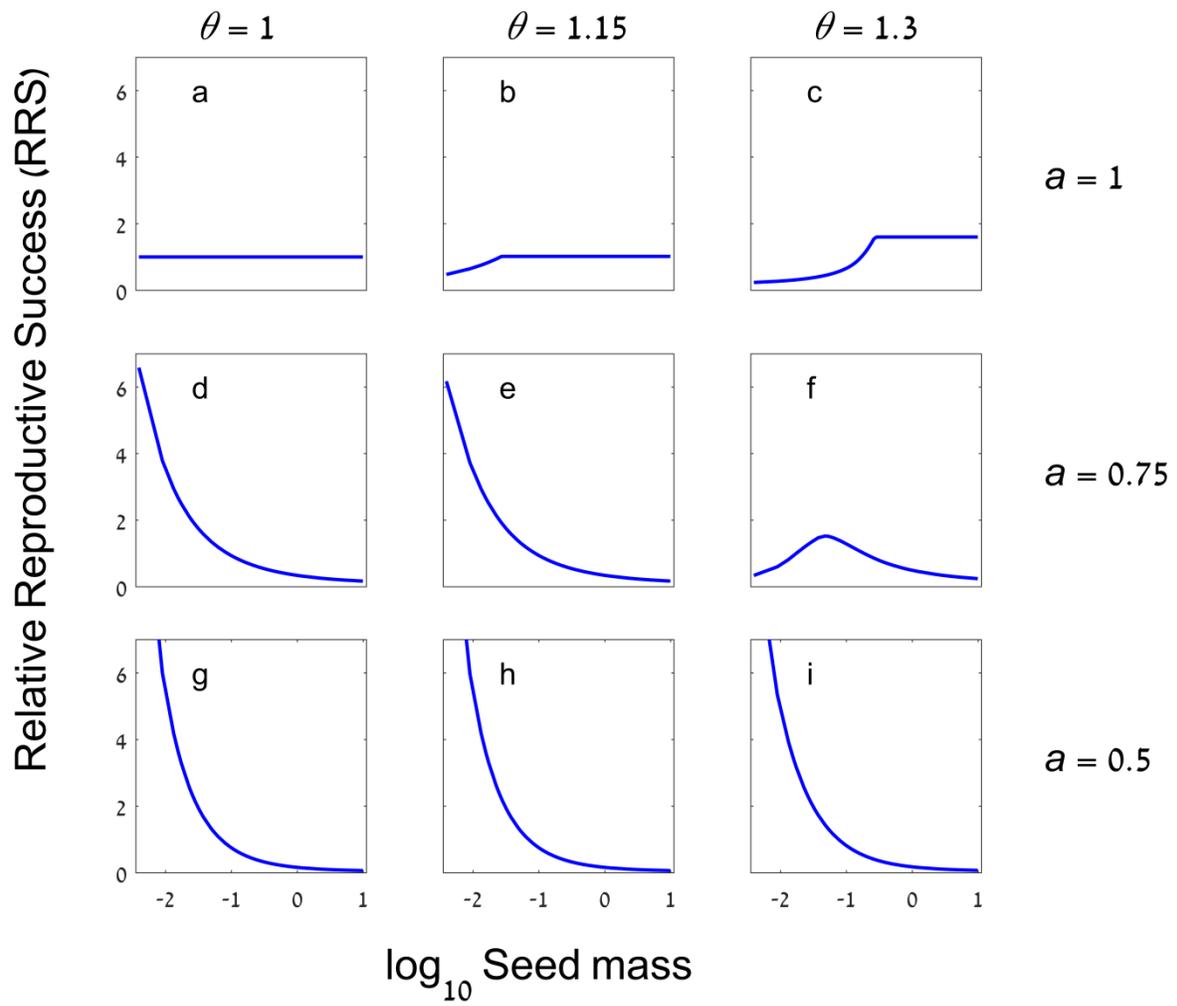

**Figure 2**

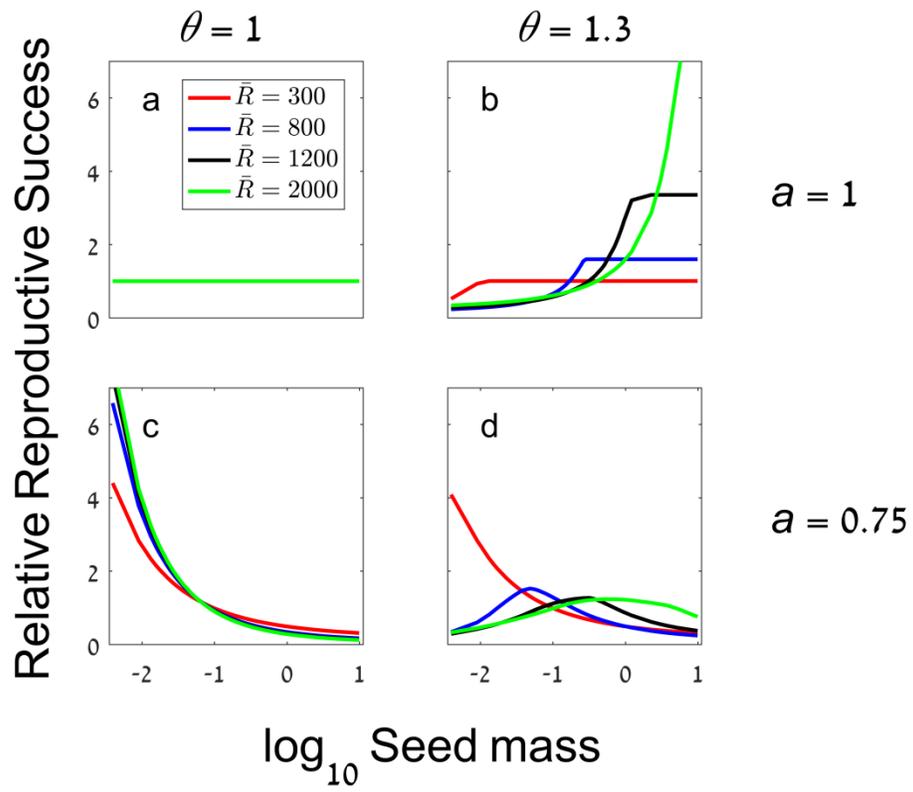

**Figure 3**

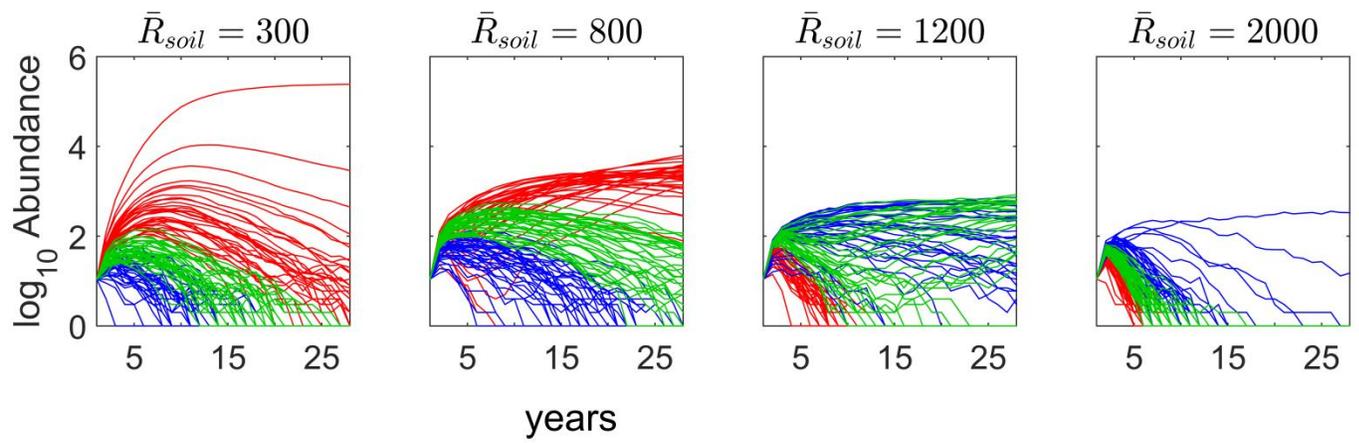

**Figure 4**

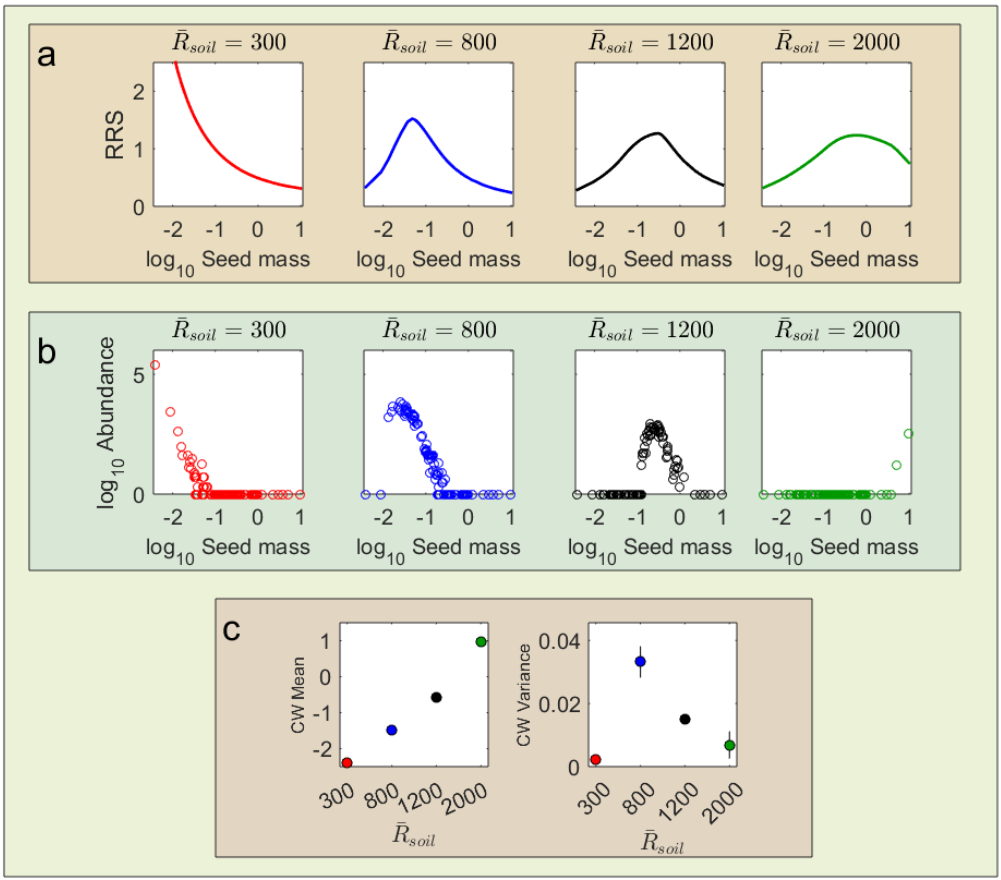